\documentclass[aps,twocolumn,showpacs,superscriptaddress]{revtex4}
\usepackage{graphicx}
\usepackage[latin1]{inputenc}
\usepackage{amsmath}
\usepackage{amssymb}

\newcommand{\nc}{\newcommand}
\nc{\bra}[1]{\langle #1|}
\nc{\ket}[1]{|#1\rangle}
\nc{\braket}[1]{\left\langle #1 \right\rangle}
\nc{\equ}[1]{\begin{eqnarray*}#1\end{eqnarray*}}
\nc{\equn}[1]{\begin{eqnarray}#1\end{eqnarray}}
\nc{\dagg}{^{\dagger}}
\nc{\conj}{^{*}}
\nc{\dx}[1]{\, \mathrm{d} {#1} \,}
\nc{\Dx}[1]{\mathcal{D} {#1} \,}
\nc{\la}{\langle}
\nc{\ra}{\rangle}
\nc{\tr}{\text{tr}}
\nc{\Tr}{\text{Tr} \,}
\nc{\e}{\text{e}}
\nc{\Id}{\mathbb{1}}
\nc{\eps}{\varepsilon}
\nc{\der}[2]{\frac{\mathrm{d} {#1}}{\mathrm{d} {#2}}}
\nc{\pder}[2]{\frac{\partial {#1}}{\partial {#2}}}
\nc{\bigO}{\mathcal{O}}

\nc{\Eq}[1]{Eq.(\ref{#1})}
\nc{\eq}[1]{Eq.(\ref{#1})}
\nc{\chap}[1]{Chapter \ref{#1}}
\nc{\Sect}[1]{Section \ref{#1}}
\nc{\sect}[1]{section \ref{#1}}
\nc{\fig}[1]{Fig.\ref{#1}}
\nc{\Fig}[1]{Figure \ref{#1}}
\nc{\tabl}[1]{Table \ref{#1}}
\nc{\app}[1]{Appendix \ref{#1}}

\nc{\eg}{\emph{e.g.} }
\nc{\ie}{\emph{i.e.} }
\nc{\etal}{et al.}

\begin{document}

\title{Dynamics of genuine multipartite correlations in open quantum systems}

\author{Arne L. Grimsmo}\email{arne.grimsmo@ntnu.no}
\affiliation{Department of Physics, 
             University of Auckland, Private Bag 92019, Auckland, New Zealand}
\affiliation{Department of Physics, 
             The Norwegian University of Science and Technology, N-7491 Trondheim, Norway}

\author{Scott Parkins}\email{s.parkins@auckland.ac.nz}
\affiliation{Department of Physics, 
             University of Auckland, Private Bag 92019, Auckland, New Zealand}

\author{Bo-Sture K. Skagerstam}\email{bo-sture.skagerstam@ntnu.no}
\affiliation{Department of Physics, 
             The Norwegian University of Science and Technology, N-7491 Trondheim, Norway}
\affiliation{Centre for Advanced Study (CAS), Drammensveien 78, N-0271 Oslo, Norway}

\date{\today}

\begin{abstract}
We propose a measure for genuine multipartite correlations suited for the study of dynamics in open quantum systems. This measure is contextual in the sense that it depends on how information is read from the environment. It is used to study an interacting collective system of atoms undergoing phase transitions as external parameters
are varied. We show that the steady state of the system can have a significant degree of genuine multipartite quantum and classical correlations, and that the proposed measure can serve as a witness of critical
behavior in quantum systems.
\end{abstract}


% 05.30.-d : Fermi-Dirac statistics (LMG?)
% 03.65.Ud: GHZ states
% 75.10.−b: Magnetic ordering

\maketitle

\emph{Introduction.---}There is a lot of interest in the characterization and quantification of correlations between quantum systems. The fact that quantum systems can be correlated in ways that classical systems can not is generally understood to be essential to emerging quantum technologies \cite{NielsenChuang00}, and thus understanding the nature of such correlations is one of the most active areas of research in quantum information theory. Recently, there has been much effort made towards quantifying correlations---quantum or classical---for \emph{multipartite} (as opposed to bipartite) systems. In particular, there exists a question of whether or not a system contains \emph{genuine} multipartite correlations, i.e., are there correlations that cannot be accounted for by any subsystem alone. A general framework, where a measure for a type of correlation was defined as the ``distance'', as measured by relative entropy, from one state with a defining property to the closest state without said property, was recently proposed by Modi et al. in \cite{Modi10}. This framework permits a unified view of the measures of relative entropy of entanglement \cite{Vedral97}, quantum discord \cite{Ollivier01}, classical correlations \cite{HendersonVedral01}, quantum mutual information and several new quantities. A powerful feature of this framework is that the suggested measures are well defined, and relations between them also hold for multipartite systems, thereby allowing for natural extensions of the above quantities beyond the bipartite case. A natural follow up to the work in \cite{Modi10} was done by Giorgi et al. in \cite{Giorgi11} where $n$-partite quantum and classical correlations were further divided into genuine and non-genuine parts. This is not the only effort in this direction however, and it is not known how to correctly measure genuine multipartite correlations. Investigations of the validity of the various measures involving the dynamical evolution of correlations in experimentally relevant situations therefore seem beneficial. This is the purpose of the work presented here.

A notorious feature of many of the proposed measures of correlations in quantum systems is that they involve difficult extremizations over operators or states. This makes calculations hard in practical applications and restricts analytical results to simple examples. Even numerically, the extremization required is often intractable. Quite generically though, the extremization is not necessary for pure states, for which the measures are greatly simplified. The best known example is for bipartite entanglement, for which all the various mixed state entanglement measures reduce to the entropy of the reduced state of one of the subsystems. This gives hope for a dynamical characterization of correlations when the dynamics themselves naturally give rise to a decomposition of the quantum state into an ensemble of pure states.  For bipartite entanglement this has been exploited to define an \emph{average} entanglement for open Markovian systems with an evolution described by a Lindblad master equation \cite{NhaCarmichael04}, and much work has been done in recent years to investigate this entanglement measure in various experimentally relevant situations \cite{Carvalho07,Vogelsberger10,Viviescas10,Mascarenhas11,Carvalho11}. We present an analogous definition for genuine multipartite correlations and show how it can be used to study the dynamical evolution of correlations in open multipartite quantum systems. We explore this new measure by applying it first to very simple three-qubit systems, before moving on to the more physically interesting case of a collective spin state of a large number of atoms that undergoes phase transitions as an external parameter is varied. For this last case, the proposed measure of genuine multipartite correlations is shown to be a good witness for the critical behavior of the system.

\emph{Average genuine multipartite correlations.---}We use the definitions of genuine $n$-partite correlations put forth in \cite{Giorgi11}. As was shown, genuine correlations can be considered to be bipartite measures for some particular bipartite cut of the system, and for a pure state $\rho$ genuine classical and quantum correlations therefore coincide, and are given simply as the entropy of the reduced state of subsystem $i$, minimized over all possible bipartitions \cite{Modi10,Giorgi11}:
\begin{align}\label{eq:corr_pure}
\mathcal{J}^{(n)}(\rho) = \mathcal{D}^{(n)}(\rho) &= \mathcal{T}^{(n)}(\rho)/2 = \min_i [S(\rho_i)] .% \\
%&(\rho \text{ pure})\nonumber,
\end{align}
Here, $\mathcal{J}^{(n)}$, $\mathcal{D}^{(n)}$ and $\mathcal{T}^{(n)}$ refer to the genuine classical, quantum, and total correlations, respectively. We note that this quantity was also considered as a multipartite entanglement measure in \cite{PopeMilburn03}. We wish to consider the dynamical evolution of these correlations for an open quantum system consisting of several sub-systems. We will consider an environment such that the Born-Markov approximation applies and that the reduced density matrix of the system, $\rho$, evolves according to a master equation in Lindblad form \cite{Breuer07} ($\hbar=1$)
\begin{align}\label{eq:master}
\dot{\rho} = -i [H,\rho] + \sum_i D[J_i]\rho.
\end{align}

Here $D[O]\rho = O \rho O\dagg - \frac{1}{2} O\dagg O \rho - \frac{1}{2} \rho O\dagg O$, and the $J_i$ are ``jump operators'' acting on the system, describing the environment in terms of various types of noise. Given a pure initial state $\rho = \ket{\psi}\bra{\psi}$, the quantum trajectory method \cite{Dalibard92,Carmichael93} evolves the state in time step $\dx{t}$ from $\ket{\psi(t)}$ to either $J_i\ket{\psi(t)}/||J_i|\psi(t)||$, with probability $p_i = \langle \psi(t)|J_i\dagg J_i|\psi(t) \rangle \dx{t}$, or to $\exp(-i H_\text{eff}t/\hbar)\ket{\psi(t)}/||\exp(-i H_\text{eff}t/\hbar)\ket{\psi(t)}||$, with probability $1-\sum_i p_i$, where $H_\text{eff} = H - i\hbar/2 \sum_i J_i\dagg J_i$. This means that the state $\ket{\psi(t)}$ evolved in this way is pure for all times $t$, and \eq{eq:corr_pure} applies. The density matrix $\rho(t)$ can be retrieved as $\rho(t) = \overline{\ket{\psi_j(t)}\bra{\psi_j(t)}}$, where each $\ket{\psi_j(t)}$ is a stochastic realization according to the quantum trajectory method, and the bar denotes an average over all possible realizations. We thus define the \emph{average genuine n-partite correlations} for an initially pure $n$-partite state with time evolution given by \eq{eq:master} as
\begin{align}\label{eq:corr_avg}
C^{(n)}_U(t) = \overline{\mathcal{J}^{n}(\ket{\psi_j(t)}\bra{\psi_j(t)}}) = \overline{\min_i S(\tr_i \ket{\psi_j(t)}\bra{\psi_j(t)})},
\end{align}
where as before $i$ runs over all bipartitions, and $\tr_i$ denotes the corresponding partial trace.

For bipartite pure states, entanglement represents all quantum correlations \cite{Modi10}, and indeed we see that \eq{eq:corr_avg} reduces to the average entanglement over trajectories defined in \cite{NhaCarmichael04} for $n=2$. The symbol $U$ here denotes the fact that this quantity is dependent on the choice of \emph{unraveling} of \eq{eq:master}, i.e., the master equation is invariant under the unitary and inhomogeneous transformations $J_i \to G_i = \sum_j u_{ij} J_j + \alpha_i$, $H \to H + 1/2i (\alpha_i\conj J_i - \alpha_i J_i\dagg)$, where $u_{ij}$ is a unitary matrix and the $\alpha_i$ are complex numbers. In other words, there is a freedom in the choice of jump operators, and different choices will in general produce different trajectories and therefore a different time evolution for the average genuine correlations.

\emph{Three-qubit systems.---}To illustrate some of the new behavior in going from the bipartite to the multipartite case, we first concentrate on the simplest example of three qubits. We restrict our attention from the vast variety of different interactions between the qubits and their environments to the case of spontaneous emission of atomic qubits with the qubits  self-Hamiltonian set to zero, $H=0$. That is, we consider unitary combinations of the jump operators $\sqrt{\gamma_A} a$, $\sqrt{\gamma_B} b$ and $\sqrt{\gamma_C} c$, where $\gamma_{A,B,C}$ are decay rates and $a = \ket{0}\bra{1}_A$ is the lowering operator on qubit $A$ and similarly for $b$ and $c$.
 
\begin{figure}
\vspace{1mm}
\includegraphics{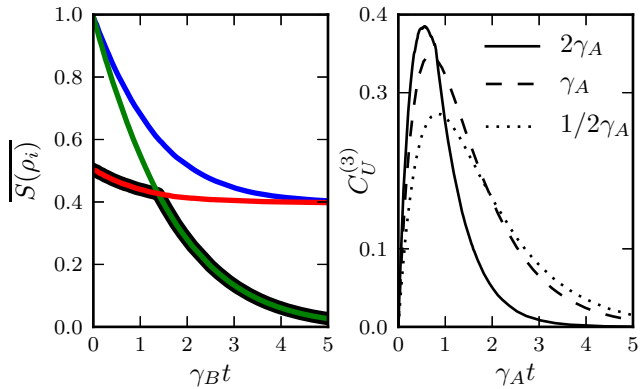}
%\vspace{0.5cm}
% %\begin{picture}(-10,0)(133,35)   %  \begin{picture}(width,height)(x-offset,y-offset)    
%\begin{picture}(100,150)(230,425)%(75,0)        
%\special{PSfile=fig1.eps
%         hscale=100  vscale=100
%         angle=0}
%\end{picture}
%\vspace{-0.5cm}
\caption{(Color online). Left: Average entropies for the three 
qubits, $\overline{S(\rho_\text{A})}$ (blue), $\overline{S(\rho_B)}$ (green) and $\overline{S(\rho_C)}$ (blue) for initial state $\ket{\psi}_1$ and with decay of qubit $B$ only. The minimum, $C^{(3)}_U$, is marked by the thick black line. Right: $C^{(3)}_U$ for the initial state $\ket{\psi}_2$ when the decay channels of qubits $A$ and $B$ 
are mixed at a beam splitter before detection, for three different $\gamma_B$.}
\label{fig:threequbits}
\end{figure}

As a first example, we consider the initial state $\ket{\psi}_1 = (2\ket{011}+2\ket{101}+1\ket{110})/\sqrt{9}$ and the case where just qubit $B$ is decaying (i.e., $\gamma_B>0$, $\gamma_A=\gamma_C=0$). The initial qubit entropies are $S(\rho_A) = S(\rho_B) = 1$ and $C^{(3)}_U = S(\rho_C) = 1/2$. Due to the decay of qubit $B$ the entropy $S(\rho_B(t))$ will eventually become smaller than $S(\rho_C(t))$, so that $C^{(3)}_U$ has a smooth behavior everywhere except at the point in time when the entropy of qubit $B$ is equal to that of qubit $C$. This is illustrated in the left panel of \fig{fig:threequbits}. This non-smooth behavior is only possible for three or more qubits and ultimately stems from the symmetrization over bipartite cuts in the definition of genuine $n$-partite correlations. 

Next we consider an example where the coupling of the system to an environment can create genuine correlations even if no such correlations are present initially. We take $\ket{\psi}_2 = 1/\sqrt{2} (\ket{010}+\ket{001})$ as our initial state. This gives the initial entropies $S(\rho_B) = S(\rho_C) = 1$ and $S(\rho_A) = 0$. Under a direct photo-detection scheme, i.e., jump operators $\sqrt{\gamma_A} a$, $\sqrt{\gamma_B} b$ and $\sqrt{\gamma_C} c$, the genuine 3-partite correlations $C^{(3)}_U$ will stay zero for all times. If two of the decay channels are combined (e.g., in a beam splitter) however, we have the possibility of genuine 3-partite correlations being created in the system from measurement back-action. We consider the unraveling $J_1 = 1/\sqrt{2} (\sqrt{\gamma_A} a + i\sqrt{\gamma_B} b)$, $J_2 = 1/\sqrt{2} (i\sqrt{\gamma_A} a + \sqrt{\gamma_B} b)$ and $J_3 = \sqrt{\gamma_C} c$. In the right panel of \fig{fig:threequbits} we show $C^{(3)}_U$ for this unraveling as a function of $\gamma_A t$ for three different $\gamma_B$ (and $\gamma_C = \gamma_A$), showing that the maximum of the genuine 3-partite correlations in fact grows with the damping of qubit $B$. Similar counter-intuitive behavior of environment-assisted entanglement that grows with the damping of a cavity mode has been shown previously for bipartite entanglement \cite{NhaCarmichael04}, but it has a different source in our case, namely that the entropy of a subsystem increases under a detection scheme where its output is mixed with another subsystem of higher entropy.

\emph{Critical behavior in collective spin systems.---}Collective spin systems are an interesting class of many-qubit systems for which states and dynamics are symmetric with regards to permutations of the qubits and can be described, e.g., in terms of Dicke states \cite{Dicke54}, which are simultaneous eigenstates of the collective angular momentum $J$ and its $z$ component $J_z$. Physically, such systems arise when the dynamics is such that excitations cannot be uniquely identified with individual qubits. 
An $N$-qubit symmetric Dicke state with $m$ excitations is given by  \cite{Stockton03}
\begin{align}
\label{eq:Dicke}
\ket{m,N} = \binom{N}{m}^{-1/2} \sum_k P_k\left(\ket{1_1,\dots,1_m,0_{m+1},\dots,0_N}\right),
\end{align}
where $\{P_k\}$ denotes the set of all distinct permutations of the spins.

Quantum phase transitions have been studied theoretically for such systems and shown to be marked by critical behavior of bipartite entanglement measures (between either pairs of spins or blocks of spins) \cite{Stockton03,Vidal04,Vidal05,MorrisonParkins08,Cui08}, and recently in the quantum correlations between pairs of spins \cite{Wang12}. We show now that this is also the case for our measure of genuine $N$-partite correlations. The specific (and topical) model we consider is the Lipkin-Meshkov-Glick (LMG) model \cite{Lipkin65}, which is described by the Hamiltonian
\begin{align}
H_\text{LMG} = -2hJ_z - 2(\lambda/N)(J_x^2 + \gamma J_y^2),
\end{align}
where the $J_x$, $J_y$ and $J_z$ are collective angular momentum operators, $h$ and $\lambda$ are effective magnetic field and spin-spin interactions strengths, respectively, and $\gamma \in [-1,1]$ is an anisotropy parameter, which we will set to zero. A dissipative version of this model was studied in \cite{MorrisonParkins08} based on the collective interaction of an ensemble of atoms with laser fields and field modes of a high-finesse optical resonator. Under appropriate conditions, the cavity modes may be adiabatically eliminated to give a master equation describing the collective atomic system alone,
\begin{align}
\label{eq:master_LMG}
\dot{\rho} =& -i[H_{\text{LMG}},\rho] + \frac{\Gamma_b}{N} D[J_+]\rho  ,
\end{align}
where $J_+ = \sum_j \ket{1_j}\bra{0_j}$ is the collective raising operator and $J_-$ its Hermitian conjugate. The quantum jumps described by $J_+$ can be associated with photo-detections of one of the cavity modes. 

\Eq{eq:master_LMG} has a steady state solution, $\rho_\text{ss}$, that exhibits both 1st and 2nd order phase transitions as the effective magnetic field strength $h$ is varied in relation to the spin-spin interaction strength $\lambda$; in particular, around $h=0$ and $h=\lambda(>0)$, respectively \cite{MorrisonParkins08}. To find the average genuine $N$-partite correlations, \eq{eq:corr_avg}, we note that a pure state of the system, evolved according to the quantum trajectory method, can be written as a linear combination of states of the form \eq{eq:Dicke}. Any such symmetric Dicke state has a Schmidt decomposition under the arbitrary bipartition $(1,2,\dots,N_1)(N_1+1,N_1+2,\dots,N)$ that can be written as \cite{Stockton03}
\begin{align}
\ket{m,N} = \sum_k \lambda_k^m \ket{k,N_1} \otimes \ket{m-k,N-N_1} ,
\end{align}
where $\lambda_k^m = \binom{N}{m}^{-1/2} \binom{N_1}{k}^{1/2} \binom{N-N_1}{m-k}^{1/2}$.
For a state $\rho = \sum_{m,n} c_m c_n\conj \ket{m,N}\bra{n,N}$ we can then find the state reduced over the first $N_1$ qubits:
\begin{align}
\rho_{N_1} \equiv& \tr_{1,2,\dots,N_1} \rho = \sum_{m,n} c_m c_n\conj \sum_k \lambda_k^m (\lambda_k^n)\conj \nonumber \\
&\times \ket{m-k,N-N_1}\bra{n-k,N-N_1}.
\end{align}
The genuine $N$-partite correlation is then the entropy of this state minimized over all possible bipartitions ($\rho$ pure):
\begin{align}
\label{eq:corr_pure_Dicke}
\mathcal{J}^{(N)}(\rho) = \min_{N_1} S(\rho_{N_1}).
\end{align}

\begin{figure}
\vspace{1mm}
\includegraphics{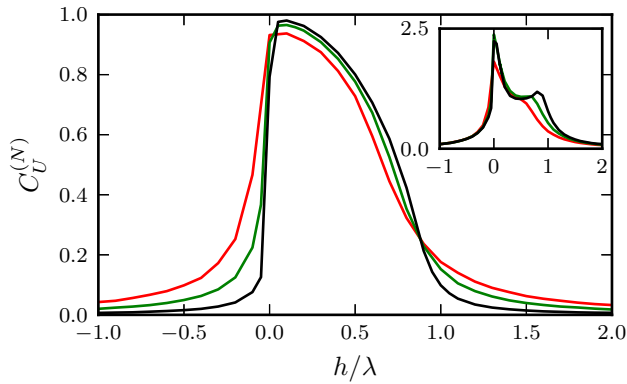}
%\vspace{0.1cm}
% %\begin{picture}(-10,0)(133,35)   %  \begin{picture}(width,height)(x-offset,y-offset)    
%\begin{picture}(100,150)(75,0)        
%\special{PSfile=fig2.eps
%         hscale=100  vscale=100
%         angle=0}
%\end{picture}
%%\vspace{1.0cm}
\caption{(Color online) Average genuine $N$-partite correlations, $C_U^{(N)}$, versus $h/\lambda$ for the dissipative LMG model with $\gamma = 0$, $\Gamma_b/\lambda = 0.2$, and $N=10$ (red), $N=25$ (green) and $N=100$ (black) qubits. In the inset we show the complementary quantity that \emph{maximizes} the entropy over all bipartitions.}
\label{fig:LMG_min}
%\vspace{30mm}
\end{figure}

In \fig{fig:LMG_min} we show numerical results for the average genuine $N$-partite correlations, \eq{eq:corr_avg}, after letting the system reach its steady state and averaging over a large number of quantum trajectories with the jump operator $\sqrt{2\Gamma_b/N} J_+$. The parameters used were $\Gamma_b = 0.2$ and $N=10$, $25$ and $100$ qubits. With increasing $N$, the behavior of $C_U^{(N)}$ points clearly to 1st and 2nd order phase transitions around $h/\lambda=0$ and $h/\lambda=1$, respectively. 
In the inset of the figure we also show the average of $\max_{N_1} S(\rho_{N_1})$. Although not a measure of genuine correlations, this quantity also shows very marked behavior around the critical points.

\begin{figure}
\vspace{1mm}
\includegraphics{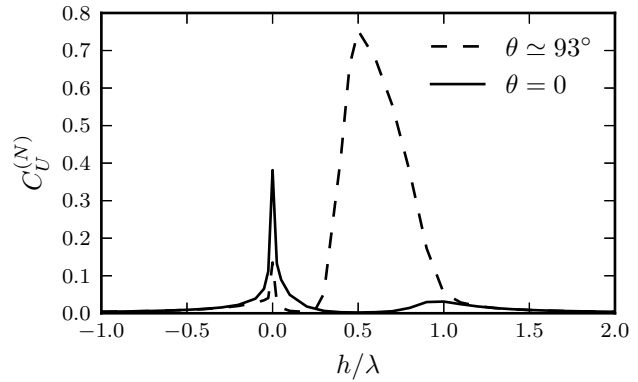}%
%\vspace{0.5cm}
% \begin{picture}(-10,0)(133,35)   %  \begin{picture}(width,height)(x-offset,y-offset)              
%\special{PSfile=fig3.eps
%         hscale=100  vscale=100
%         angle=0}
%\end{picture}
%\vspace{0.5cm}
\caption{Average genuine $N$-partite correlations versus $h/\lambda$ for $N=100$ and homodyne detection with two different detection angles. The angle $\theta = 92.8^\circ$ maximizes the correlations at $h/\lambda = 0.5$.}
\label{fig:LMG_min_homodyne}
\end{figure}

We can also study alternative, diffusive unravelings of the master equation \eq{eq:master_LMG} by considering, e.g., homodyne detection, in which the output cavity field is combined at a beam-splitter with a strong laser field before detection. In the present context, this corresponds to measurement of the spin quadrature $X_\theta = \e^{i\theta} J_+ + \e^{-i\theta}J_-$. The evolution of a quantum trajectory $\ket{\psi_j(t)}$ can then be given as a continuous stochastic differential equation \cite{Breuer07,WisemanMilburn93},
\begin{align}
&\dx{\ket{\psi_j(t)}} = -iH\ket{\psi_j(t)} \dx{t} \nonumber\\
&+ \frac{\Gamma_b}{2N} \bigg\{ \braket{X_\theta}\e^{-i\theta}J_+ - J_- J_+ - \frac{1}{4}\braket{X_\theta}^2 \bigg\} \ket{\psi_j(t)} \dx{t}\nonumber\\
&+ \sqrt{\frac{\Gamma_b}{N}} \bigg\{\e^{-i\theta}J_+ - \frac{1}{2}\braket{X_\theta} \bigg\} \ket{\psi_j(t)} \dx{W(t)},
\end{align}
where $\dx{W(t)}$ is a real Wiener increment. 
Results for $C_U^{(N)}$ with this unraveling are shown in \fig{fig:LMG_min_homodyne} for $N=100$ and two different detection angles $\theta$. Considerable sensitivity of $C_U^{(N)}$ to $\theta$ is apparent in the region between $h/\lambda=0$ and $1$. In this region, the system state during a quantum trajectory can approach a superposition of $J_x$ eigenstates of opposite signs (note that the spin $Q$-function computed from $\rho_{\rm ss}$ has a two-peaked structure along the $J_x$-axis \cite{MorrisonParkins08_2}), provided the homodyne measurement is unable to resolve these eigenstates, i.e., provided the phase $\theta\simeq 90^\circ$, corresponding to measurement of $J_y$. Such a superposition of $J_x$ eigenstates is strongly correlated, giving rise to a large value of $C_U^{(N)}$, while a single $J_x$ eigenstate, towards which the state is projected by a measurement with $\theta=0^\circ$, has no genuine correlations and hence a negligible value of $C_U^{(N)}$ over much of the region between $h/\lambda=0$ and $1$. We note that the detection angle that gives the maximum value of $C_U^{(N)}$ varies with both $h/\lambda$ and $N$, and that the sensitivity increases with $N$.

We can of course write the steady state solution of \eq{eq:master_LMG} as a weighted ensemble of pure states in infinitely many ways, $\rho_\text{ss} = \sum_k p_k \ket{\psi_k}\bra{\psi_k}$, but only some of these are so-called \emph{physically realizable ensembles} \cite{WisemanVaccaro01}, i.e., there exists some continuous measurement of the environment such that the state is collapsed to $\ket{\psi_k}$ a proportion of the time $p_k$. This has been referred to as the \emph{preferred ensemble  fact}. These ensembles are just the ones attainable through the quantum trajectory method, which gives a physical interpretation to the average genuine correlations studied above.

\emph{Conclusions.---}To define a measure of genuine multipartite correlations suitable for the dynamical study of open quantum systems, we have exploited the fact that in many experimentally relevant situations the dynamics naturally give rise to a decomposition of the state into an ensemble of pure states. Even though the decomposition might not have any clear information theoretic interpretation, it has the advantage of arising from a well-defined physical process, as well as being well suited for computation through computer simulations. In a sense this measure generalizes the measure for average bipartite entanglement over quantum trajectories defined previously \cite{NhaCarmichael04}, and we have explored how going from bipartite to multipartite systems can give new exotic behavior in the dynamical evolution of average genuine correlations. Just as for the bipartite case we have seen that the measure can be sensitive with respect to the choice of unraveling, $U$, which leaves many questions open with respect to the significance of different detection schemes. Finally, we have shown numerical evidence that the proposed measure of genuine correlations can serve as a witness for critical behavior in open quantum systems.

%\begin{acknowledgments}
%\end{acknowledgments}

%\bibliography{arne2}

\begin{thebibliography}{23}
\expandafter\ifx\csname natexlab\endcsname\relax\def\natexlab#1{#1}\fi
\expandafter\ifx\csname bibnamefont\endcsname\relax
  \def\bibnamefont#1{#1}\fi
\expandafter\ifx\csname bibfnamefont\endcsname\relax
  \def\bibfnamefont#1{#1}\fi
\expandafter\ifx\csname citenamefont\endcsname\relax
  \def\citenamefont#1{#1}\fi
\expandafter\ifx\csname url\endcsname\relax
  \def\url#1{\texttt{#1}}\fi
\expandafter\ifx\csname urlprefix\endcsname\relax\def\urlprefix{URL }\fi
\providecommand{\bibinfo}[2]{#2}
\providecommand{\eprint}[2][]{\url{#2}}

\bibitem[{\citenamefont{Nielsen and Chuang}(2000)}]{NielsenChuang00}
\bibinfo{author}{\bibfnamefont{M.}~\bibnamefont{Nielsen}} \bibnamefont{and}
  \bibinfo{author}{\bibfnamefont{I.}~\bibnamefont{Chuang}},
  \emph{\bibinfo{title}{Quantum Computation and Quantum Information}}
  (\bibinfo{publisher}{Cambridge University Press}, \bibinfo{year}{2000}).

\bibitem[{\citenamefont{Modi et~al.}(2010)\citenamefont{Modi, Paterek, Son,
  Vedral, and Williamson}}]{Modi10}
\bibinfo{author}{\bibfnamefont{K.}~\bibnamefont{Modi}},
  \bibinfo{author}{\bibfnamefont{T.}~\bibnamefont{Paterek}},
  \bibinfo{author}{\bibfnamefont{W.}~\bibnamefont{Son}},
  \bibinfo{author}{\bibfnamefont{V.}~\bibnamefont{Vedral}}, \bibnamefont{and}
  \bibinfo{author}{\bibfnamefont{M.}~\bibnamefont{Williamson}},
  \bibinfo{journal}{Phys. Rev. Lett.} \textbf{\bibinfo{volume}{104}},
  \bibinfo{pages}{080501} (\bibinfo{year}{2010}).

\bibitem[{\citenamefont{Vedral et~al.}(1997)\citenamefont{Vedral, Plenio,
  Rippin, and Knight}}]{Vedral97}
\bibinfo{author}{\bibfnamefont{V.}~\bibnamefont{Vedral}},
  \bibinfo{author}{\bibfnamefont{M.~B.} \bibnamefont{Plenio}},
  \bibinfo{author}{\bibfnamefont{M.~A.} \bibnamefont{Rippin}},
  \bibnamefont{and} \bibinfo{author}{\bibfnamefont{P.~L.}
  \bibnamefont{Knight}}, \bibinfo{journal}{Phys. Rev. Lett.}
  \textbf{\bibinfo{volume}{78}}, \bibinfo{pages}{2275} (\bibinfo{year}{1997}).

\bibitem[{\citenamefont{Ollivier and Zurek}(2001)}]{Ollivier01}
\bibinfo{author}{\bibfnamefont{H.}~\bibnamefont{Ollivier}} \bibnamefont{and}
  \bibinfo{author}{\bibfnamefont{W.~H.} \bibnamefont{Zurek}},
  \bibinfo{journal}{Phys. Rev. Lett.} \textbf{\bibinfo{volume}{88}},
  \bibinfo{pages}{017901} (\bibinfo{year}{2001}).

\bibitem[{\citenamefont{Henderson and Vedral}(2001)}]{HendersonVedral01}
\bibinfo{author}{\bibfnamefont{L.}~\bibnamefont{Henderson}} \bibnamefont{and}
  \bibinfo{author}{\bibfnamefont{V.}~\bibnamefont{Vedral}},
  \bibinfo{journal}{J. Phys. A: Math. Gen.}
  \textbf{\bibinfo{volume}{34}}, \bibinfo{pages}{6899} (\bibinfo{year}{2001}).

\bibitem[{\citenamefont{Giorgi et~al.}(2011)\citenamefont{Giorgi, Bellomo,
  Galve, and Zambrini}}]{Giorgi11}
\bibinfo{author}{\bibfnamefont{G.~L.} \bibnamefont{Giorgi}},
  \bibinfo{author}{\bibfnamefont{B.}~\bibnamefont{Bellomo}},
  \bibinfo{author}{\bibfnamefont{F.}~\bibnamefont{Galve}}, \bibnamefont{and}
  \bibinfo{author}{\bibfnamefont{R.}~\bibnamefont{Zambrini}},
  \bibinfo{journal}{Phys. Rev. Lett.} \textbf{\bibinfo{volume}{107}},
  \bibinfo{pages}{190501} (\bibinfo{year}{2011}).

\bibitem[{\citenamefont{Nha and Carmichael}(2004)}]{NhaCarmichael04}
\bibinfo{author}{\bibfnamefont{H.}~\bibnamefont{Nha}} \bibnamefont{and}
  \bibinfo{author}{\bibfnamefont{H.~J.} \bibnamefont{Carmichael}},
  \bibinfo{journal}{Phys. Rev. Lett.} \textbf{\bibinfo{volume}{93}},
  \bibinfo{pages}{120408} (\bibinfo{year}{2004}).

\bibitem[{\citenamefont{Carvalho et~al.}(2007)\citenamefont{Carvalho, Busse,
  Brodier, Viviescas, and Buchleitner}}]{Carvalho07}
\bibinfo{author}{\bibfnamefont{A.~R.~R.} \bibnamefont{Carvalho}},
  \bibinfo{author}{\bibfnamefont{M.}~\bibnamefont{Busse}},
  \bibinfo{author}{\bibfnamefont{O.}~\bibnamefont{Brodier}},
  \bibinfo{author}{\bibfnamefont{C.}~\bibnamefont{Viviescas}},
  \bibnamefont{and}
  \bibinfo{author}{\bibfnamefont{A.}~\bibnamefont{Buchleitner}},
  \bibinfo{journal}{Phys. Rev. Lett.} \textbf{\bibinfo{volume}{98}},
  \bibinfo{pages}{190501} (\bibinfo{year}{2007}).

\bibitem[{\citenamefont{Vogelsberger and Spehner}(2010)}]{Vogelsberger10}
\bibinfo{author}{\bibfnamefont{S.}~\bibnamefont{Vogelsberger}}
  \bibnamefont{and} \bibinfo{author}{\bibfnamefont{D.}~\bibnamefont{Spehner}},
  \bibinfo{journal}{Phys. Rev. A} \textbf{\bibinfo{volume}{82}},
  \bibinfo{pages}{052327} (\bibinfo{year}{2010}).

\bibitem[{\citenamefont{Viviescas et~al.}(2010)\citenamefont{Viviescas,
  Guevara, Carvalho, Busse, and Buchleitner}}]{Viviescas10}
\bibinfo{author}{\bibfnamefont{C.}~\bibnamefont{Viviescas}},
  \bibinfo{author}{\bibfnamefont{I.}~\bibnamefont{Guevara}},
  \bibinfo{author}{\bibfnamefont{A.~R.~R.} \bibnamefont{Carvalho}},
  \bibinfo{author}{\bibfnamefont{M.}~\bibnamefont{Busse}}, \bibnamefont{and}
  \bibinfo{author}{\bibfnamefont{A.}~\bibnamefont{Buchleitner}},
  \bibinfo{journal}{Phys. Rev. Lett.} \textbf{\bibinfo{volume}{105}},
  \bibinfo{pages}{210502} (\bibinfo{year}{2010}).

\bibitem[{\citenamefont{Mascarenhas et~al.}(2011)\citenamefont{Mascarenhas,
  Cavalcanti, Vedral, and Santos}}]{Mascarenhas11}
\bibinfo{author}{\bibfnamefont{E.}~\bibnamefont{Mascarenhas}},
  \bibinfo{author}{\bibfnamefont{D.}~\bibnamefont{Cavalcanti}},
  \bibinfo{author}{\bibfnamefont{V.}~\bibnamefont{Vedral}}, \bibnamefont{and}
  \bibinfo{author}{\bibfnamefont{M.~F.} \bibnamefont{Santos}},
  \bibinfo{journal}{Phys. Rev. A} \textbf{\bibinfo{volume}{83}},
  \bibinfo{pages}{022311} (\bibinfo{year}{2011}).

\bibitem[{\citenamefont{Carvalho and Santos}(2011)}]{Carvalho11}
\bibinfo{author}{\bibfnamefont{A.~R.~R.} \bibnamefont{Carvalho}}
  \bibnamefont{and} \bibinfo{author}{\bibfnamefont{M.~F.}
  \bibnamefont{Santos}}, \bibinfo{journal}{New Journal of Physics}
  \textbf{\bibinfo{volume}{13}}, \bibinfo{pages}{013010}
  (\bibinfo{year}{2011}).

\bibitem[{\citenamefont{Pope and Milburn}(2003)}]{PopeMilburn03}
\bibinfo{author}{\bibfnamefont{D.~T.} \bibnamefont{Pope}} \bibnamefont{and}
  \bibinfo{author}{\bibfnamefont{G.~J.} \bibnamefont{Milburn}},
  \bibinfo{journal}{Phys. Rev. A} \textbf{\bibinfo{volume}{67}},
  \bibinfo{pages}{052107} (\bibinfo{year}{2003}).

\bibitem[{\citenamefont{Breuer and Petruccione}(2007)}]{Breuer07}
\bibinfo{author}{\bibfnamefont{H.}~\bibnamefont{Breuer}} \bibnamefont{and}
  \bibinfo{author}{\bibfnamefont{F.}~\bibnamefont{Petruccione}},
  \emph{\bibinfo{title}{The Theory of Open Quantum Systems}}
  (\bibinfo{publisher}{Oxford University Press}, \bibinfo{year}{2007}).

\bibitem[{\citenamefont{Dalibard et~al.}(1992)\citenamefont{Dalibard, Castin,
  and M\o{}lmer}}]{Dalibard92}
\bibinfo{author}{\bibfnamefont{J.}~\bibnamefont{Dalibard}},
  \bibinfo{author}{\bibfnamefont{Y.}~\bibnamefont{Castin}}, \bibnamefont{and}
  \bibinfo{author}{\bibfnamefont{K.}~\bibnamefont{M\o{}lmer}},
  \bibinfo{journal}{Phys. Rev. Lett.} \textbf{\bibinfo{volume}{68}},
  \bibinfo{pages}{580} (\bibinfo{year}{1992}).

\bibitem[{\citenamefont{Carmichael}(1993)}]{Carmichael93}
\bibinfo{author}{\bibfnamefont{H.}~\bibnamefont{Carmichael}},
  \emph{\bibinfo{title}{An Open Systems Approach to Quantum Optics}}
  (\bibinfo{publisher}{Springer}, \bibinfo{year}{1993}).

\bibitem[{\citenamefont{Dicke}(1954)}]{Dicke54}
\bibinfo{author}{\bibfnamefont{R.~H.} \bibnamefont{Dicke}},
  \bibinfo{journal}{Phys. Rev.} \textbf{\bibinfo{volume}{93}},
  \bibinfo{pages}{99} (\bibinfo{year}{1954}).

\bibitem[{\citenamefont{Stockton et~al.}(2003)\citenamefont{Stockton, Geremia,
  Doherty, and Mabuchi}}]{Stockton03}
\bibinfo{author}{\bibfnamefont{J.~K.} \bibnamefont{Stockton}},
  \bibinfo{author}{\bibfnamefont{J.~M.} \bibnamefont{Geremia}},
  \bibinfo{author}{\bibfnamefont{A.~C.} \bibnamefont{Doherty}},
  \bibnamefont{and} \bibinfo{author}{\bibfnamefont{H.}~\bibnamefont{Mabuchi}},
  \bibinfo{journal}{Phys. Rev. A} \textbf{\bibinfo{volume}{67}},
  \bibinfo{pages}{022112} (\bibinfo{year}{2003}).

\bibitem[{\citenamefont{Vidal et~al.}(2004)\citenamefont{Vidal, Palacios, and
  Aslangul}}]{Vidal04}
\bibinfo{author}{\bibfnamefont{J.}~\bibnamefont{Vidal}},
  \bibinfo{author}{\bibfnamefont{G.}~\bibnamefont{Palacios}}, \bibnamefont{and}
  \bibinfo{author}{\bibfnamefont{C.}~\bibnamefont{Aslangul}},
  \bibinfo{journal}{Phys. Rev. A} \textbf{\bibinfo{volume}{70}},
  \bibinfo{pages}{062304} (\bibinfo{year}{2004}).

\bibitem[{\citenamefont{Latorre et~al.}(2005)\citenamefont{Latorre, Or\'us,
  Rico, and Vidal}}]{Vidal05}
\bibinfo{author}{\bibfnamefont{J.~I.} \bibnamefont{Latorre}},
  \bibinfo{author}{\bibfnamefont{R.}~\bibnamefont{Or\'us}},
  \bibinfo{author}{\bibfnamefont{E.}~\bibnamefont{Rico}}, \bibnamefont{and}
  \bibinfo{author}{\bibfnamefont{J.}~\bibnamefont{Vidal}},
  \bibinfo{journal}{Phys. Rev. A} \textbf{\bibinfo{volume}{71}},
  \bibinfo{pages}{064101} (\bibinfo{year}{2005}).

\bibitem[{\citenamefont{Morrison and
  Parkins}(2008{\natexlab{a}})}]{MorrisonParkins08}
\bibinfo{author}{\bibfnamefont{S.}~\bibnamefont{Morrison}} \bibnamefont{and}
  \bibinfo{author}{\bibfnamefont{A.~S.} \bibnamefont{Parkins}},
  \bibinfo{journal}{Phys. Rev. Lett.} \textbf{\bibinfo{volume}{100}},
  \bibinfo{pages}{040403} (\bibinfo{year}{2008}{\natexlab{a}}).

\bibitem[{\citenamefont{Cui}(2008)}]{Cui08}
\bibinfo{author}{\bibfnamefont{H.~T.} \bibnamefont{Cui}},
  \bibinfo{journal}{Phys. Rev. A} \textbf{\bibinfo{volume}{77}},
  \bibinfo{pages}{052105} (\bibinfo{year}{2008}).

\bibitem[{\citenamefont{Wang et~al.}(2012)\citenamefont{Wang, Zhang, and
  Chen}}]{Wang12}
\bibinfo{author}{\bibfnamefont{C.}~\bibnamefont{Wang}},
  \bibinfo{author}{\bibfnamefont{Y.-Y.} \bibnamefont{Zhang}}, \bibnamefont{and}
  \bibinfo{author}{\bibfnamefont{Q.-H.} \bibnamefont{Chen}},
  \bibinfo{journal}{Phys. Rev. A} \textbf{\bibinfo{volume}{85}},
  \bibinfo{pages}{052112} (\bibinfo{year}{2012}).

\bibitem[{\citenamefont{Lipkin et~al.}(1965)\citenamefont{Lipkin, Meshkov, and
  Glick}}]{Lipkin65}
\bibinfo{author}{\bibfnamefont{H.}~\bibnamefont{Lipkin}},
  \bibinfo{author}{\bibfnamefont{N.}~\bibnamefont{Meshkov}}, \bibnamefont{and}
  \bibinfo{author}{\bibfnamefont{A.}~\bibnamefont{Glick}},
  \bibinfo{journal}{Nuclear Physics} \textbf{\bibinfo{volume}{62}},
  \bibinfo{pages}{188 } (\bibinfo{year}{1965}).

\bibitem[{\citenamefont{Wiseman and Milburn}(1993)}]{WisemanMilburn93}
\bibinfo{author}{\bibfnamefont{H.~M.} \bibnamefont{Wiseman}} \bibnamefont{and}
  \bibinfo{author}{\bibfnamefont{G.~J.} \bibnamefont{Milburn}},
  \bibinfo{journal}{Phys. Rev. A} \textbf{\bibinfo{volume}{47}},
  \bibinfo{pages}{642} (\bibinfo{year}{1993}).

\bibitem[{\citenamefont{Morrison and
  Parkins}(2008{\natexlab{b}})}]{MorrisonParkins08_2}
\bibinfo{author}{\bibfnamefont{S.}~\bibnamefont{Morrison}} \bibnamefont{and}
  \bibinfo{author}{\bibfnamefont{A.~S.} \bibnamefont{Parkins}},
  \bibinfo{journal}{Phys. Rev. A} \textbf{\bibinfo{volume}{77}},
  \bibinfo{pages}{043810} (\bibinfo{year}{2008}{\natexlab{b}}).

\bibitem[{\citenamefont{Wiseman and Vaccaro}(2001)}]{WisemanVaccaro01}
\bibinfo{author}{\bibfnamefont{H.~M.} \bibnamefont{Wiseman}} \bibnamefont{and}
  \bibinfo{author}{\bibfnamefont{J.~A.} \bibnamefont{Vaccaro}},
  \bibinfo{journal}{Phys. Rev. Lett.} \textbf{\bibinfo{volume}{87}},
  \bibinfo{pages}{240402} (\bibinfo{year}{2001}).

\end{thebibliography}

\end{document}